\pdfoutput=1

\documentclass[11pt]{article}
\usepackage{graphicx}
\graphicspath{{figs/}}
\usepackage[final]{acl}

\usepackage{times}
\usepackage{latexsym}
\usepackage{enumitem}
\usepackage{colortbl}
\usepackage{xcolor}
\usepackage{array}
\usepackage{filecontents}
\usepackage{tabularx}
\usepackage{threeparttable}
\usepackage{adjustbox}
\usepackage{hyperref}
\usepackage{pgfplots}
\usepackage{amssymb}
\usepackage{amsfonts}
\usepackage{url}
\usepackage{longtable}
\usepackage{rotating}
\usepackage{multirow}
\usepackage{mathrsfs}
\usepackage{tcolorbox}
\usepackage{subfigure}
\usepackage{enumitem}
\usepackage{flushend}
\usepackage{bbding}
\usepackage[toc,page]{appendix}

\definecolor{tblue}{RGB}{31,119,180}
\definecolor{torange}{RGB}{255,127,14}
\definecolor{tgreen}{RGB}{44,160,44}
\definecolor{tred}{RGB}{214,39,40}
\definecolor{tpurple}{RGB}{148,103,189}

\newcommand{\ie}{\textit{i}.\textit{e}.}
\newcommand{\eg}{\textit{e}.\textit{g}.}

\usepackage[T1]{fontenc}

\usepackage[utf8]{inputenc}

\usepackage{microtype}
\usepackage{balance}
\usepackage{inconsolata}

\usepackage{multirow}
\usepackage{subfigure}
\usepackage{amsmath}
\usepackage{booktabs}
\usepackage{amssymb}

\def\model{EasyRec}

\title{EasyRec: Simple yet Effective Language Models for Recommendation}

\author{Xubin Ren \\
  The University of Hong Kong \\
  \texttt{xubinrencs@gmail.com} \\\And
  Chao Huang\thanks{Chao Huang is the Corresponding Author.} \\
  The University of Hong Kong \\
  \texttt{chaohuang75@gmail.com} \\}

\begin{document}
\maketitle
\begin{abstract}
Deep neural networks have emerged as a powerful technique for learning representations from user-item interaction data in collaborative filtering (CF) for recommender systems. However, many existing methods heavily rely on unique user and item IDs, which restricts their performance in zero-shot learning scenarios. Inspired by the success of language models (LMs) and their robust generalization capabilities, we pose the question: How can we leverage language models to enhance recommender systems? We propose EasyRec, an effective approach that integrates text-based semantic understanding with collaborative signals. EasyRec employs a text-behavior alignment framework that combines contrastive learning with collaborative language model tuning. This ensures strong alignment between text-enhanced semantic representations and collaborative behavior information. Extensive evaluations across diverse datasets show EasyRec significantly outperforms state-of-the-art models, particularly in text-based zero-shot recommendation. EasyRec functions as a plug-and-play component that integrates seamlessly into collaborative filtering frameworks. This empowers existing systems with improved performance and adaptability to user preferences. Implementation codes are publicly available at: {\color{blue}\url{https://github.com/HKUDS/EasyRec}}.
\end{abstract}

\section{Introduction}
\label{sec:intro}

Deep learning has established itself as a highly promising solution for capturing user preferences in online recommender systems~\cite{zhang2019deep,yang2022hrcf,zhang2022geometric}. This approach harnesses deep neural networks to learn rich user and item representations by analyzing complex user-item interaction patterns. Consequently, recommendation algorithms can accurately infer user preferences and deliver personalized recommendations~\cite{xu2023causal,sun2021hgcf}.

Recent advancements in enhancing recommender systems through neural network-powered collaborative filtering frameworks, particularly graph neural networks (GNNs)~\cite{wang2019neural,he2020lightgcn,xia2022hypergraph}, have leveraged the inherent graph structure in data to capture high-order relationships among users and items. Notable GNN-based approaches, such as NGCF~\cite{wang2019neural} and LightGCN~\cite{he2020lightgcn}, demonstrate impressive performance via recursive message passing mechanisms. However, data scarcity remains a significant challenge, hindering deep collaborative filtering models from accurately learning user/item representations, especially with sparse interaction data~\cite{lin2021task,wei2021contrastive,hao2021pre}. To address this, recent studies have explored self-supervised learning for effective data augmentation. For instance, contrastive methods like SGL~\cite{wu2021self} and NCL~\cite{lin2022improving} utilize graph contrastive learning, while generative approaches like AutoCF~\cite{xia2023automated} employ masked autoencoding to reconstruct interaction structures.

Recent advances in self-supervised learning show promise in alleviating data scarcity in collaborative filtering models. However, these methods encounter a significant limitation~\cite{yuan2023go}. They rely heavily on unique identities (IDs) to represent users and items. This reliance presents major challenges for practical recommenders that must handle data from diverse domains or time periods effectively. Existing ID-based models struggle to adapt to changes in user and item identity tokens. This problem becomes particularly severe in \textbf{zero-shot recommendation} scenarios~\cite{ZESRec, UniSRec}. In these cases, training data lacks overlap with deployment data regarding users and items. This limitation hinders their ability to generalize effectively towards foundational recommender systems. Cross-domain recommendation methods~\cite{xie2022contrastive,cao2022contrastive} attempt to leverage knowledge across multiple domains. However, they often make restrictive assumptions about user overlap. These approaches assume that users from different domains belong to the same set~\cite{dacrema2012design,xie2022contrastive}. This assumption significantly limits their flexibility and generalization capabilities. Consequently, existing methods struggle to provide accurate recommendations to diverse user populations. They face particular challenges when operating across varying domains and contexts. \\\vspace{-0.08in}

\noindent \textbf{Language Models as Zero-Shot Recommenders}. The challenges discussed earlier highlight a critical need in recommendation systems. This study aims to introduce a recommender system that functions as a zero-shot learner. The system should possess robust generalization capabilities and adapt to new recommendation data seamlessly. To accomplish this objective, we propose integrating language models with collaborative relation modeling. This integration forms an effective text embedder-\model\ that is both lightweight and effective. Our approach seamlessly combines text-based semantic encoding with high-order collaborative signals. This combination results in a recommender system with strong generalization ability. The system leverages rich semantic understanding while capturing collaborative patterns from user-item interactions.

Recent research has explored leveraging large language models (LLMs) to enhance recommender systems. Existing approaches broadly fall into two categories. The first category uses LLMs for data augmentation (\eg, RLMRec~\cite{RLMRec}, AlterRec~\cite{li2024enhancing}). These methods encode textual information to complement collaborative filtering. While this combines LLM and collaborative strengths, these methods remain ID-based and struggle to generalize. The second approach utilizes LLMs to directly generate user-item interaction predictions (\eg, LLaRA~\cite{liao2024llara}, CoLLM~\cite{zhang2023collm}). However, such LLM-based recommenders suffer from poor efficiency, requiring approximately one second per prediction. This renders them impractical for large-scale recommendation tasks. These challenges highlight the need for efficient, scalable solutions that integrate semantic understanding with collaborative strengths for zero-shot recommendation.

Our model demonstrates superior performance compared to state-of-the-art language models, as illustrated in Figure~\ref{fig:scale_param}. This performance advantage is achieved within a cost-efficient parameter space of 100 to 400 million parameters. The computational cost is approximately 0.01 seconds per prediction. Notably, our model exhibits the scaling law phenomenon. Performance continually improves as parameter size increases. This contrasts with existing approaches that suffer from poor efficiency. Our model is designed to be highly scalable and practical for large-scale recommendation tasks. The computational efficiency represents significant advancement over current methods. While existing LLM-based recommenders require substantial inference time, our approach maintains both accuracy and speed for real-world deployment. In summary, this work makes the following contributions: \vspace{-0.1in}

\begin{figure}[t]
    \centering
    \includegraphics[width=1.05\columnwidth]{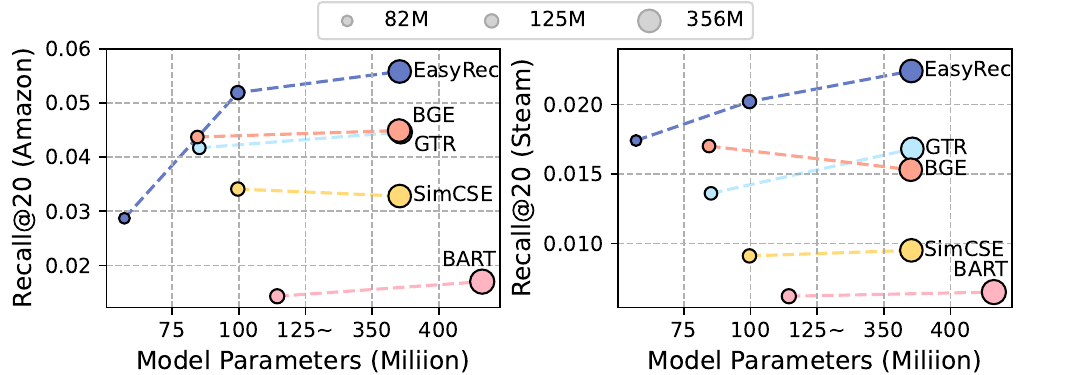}
    \vspace{-0.15in}
    \caption{\model\ outperforms state-of-the-art language models in text-based zero-shot recommendation.}
    \vspace{-0.15in}
    \label{fig:scale_param}
\end{figure}

\begin{figure*}[t]
    \centering
    \includegraphics[width=0.98\textwidth]{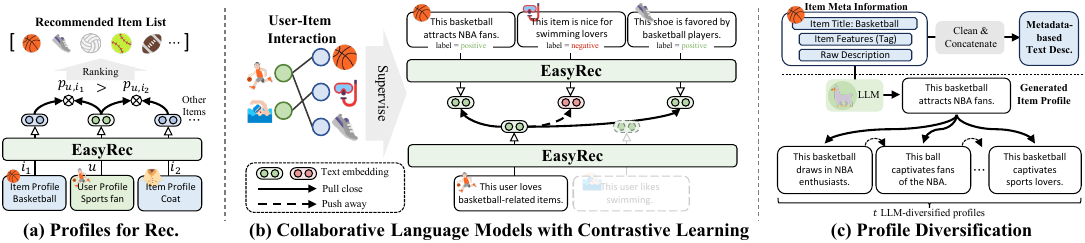}
    \vspace{-0.1in}
    \caption{The overall framework of our proposed collaborative information-guided language model \model.}
    \vspace{-0.1in}
    \label{fig:framework}
\end{figure*}

\begin{itemize}[leftmargin=*]

\item \textbf{Motivation}. The primary objective of this study is to introduce a novel recommender system built upon language models. This system functions as a zero-shot learner with exceptional adaptability to diverse recommendation data. The new \model\ exhibits robust generalization capabilities across different domains and contexts. \vspace{-0.1in}

\item \textbf{Methodology}. We propose a novel contrastive learning-powered collaborative language modeling approach. This method aligns text-based semantic encoding with collaborative signals from user behavior. The system captures both semantic representations of users and items. It also learns underlying behavioral patterns and interactions within the recommendation data. \vspace{-0.1in}

\item \textbf{Zero-Shot Recommendation Capacity}. The \model\ is extensively evaluated through rigorous experiments as a text-based zero-shot recommender system. Performance comparisons reveal consistent and significant advantages over baseline methods. The model excels in both recommendation accuracy and generalization capabilities. Furthermore, the study demonstrates remarkable potential in adapting dynamic user profiles. These profiles are highly adaptive to time-evolving user preferences. \vspace{-0.1in}

\item \textbf{Existing Recommender Enhancement}. Our proposed \model\ integrates seamlessly as a lightweight, plug-and-play component with state-of-the-art collaborative filtering models. The lightweight and modular design represents a key strength of our approach. This design facilitates adoption of our novel recommendation paradigm across diverse use cases. The integration enhances existing recommendation methods without significant computational overhead.

\end{itemize}

\section{Preliminaries}
\label{sec:preli}
In recommender systems, we have a set of users $\mathcal{U}$ and items $\mathcal{I}$, along with their interactions (e.g., clicks, purchases). For each user $u \in \mathcal{U}$, let $\mathcal{N}_u$ be the set of items interacted with. For each item $i \in \mathcal{I}$, let $\mathcal{N}_i$ be the set of users who have interacted with it. These interactions can be represented by an interaction matrix $\mathcal{A}^{|\mathcal{U}|\times|\mathcal{I}|}$. Here, $\mathcal{A}^{u, i}$ is 1 if user $u$ has interacted with item $i$, and 0 otherwise. The goal of recommendation is to predict the preference score $p_{u,i}$ of future interactions between user $u$ and item $i$. This score can be used to generate recommendations based on individual preferences. \\\vspace{-0.1in}

\noindent \textbf{Text-based Zero-Shot Recommendation} is crucial in recommender systems. It serves as a pathway toward foundation models. Textual descriptions, such as product titles and user profiles, provide valuable information. These descriptions enable effective recommendations across various datasets. This approach overcomes the limitations of traditional collaborative filtering methods. It offers significant advantage in generalization over the ID-based paradigm. This capability extends to scenarios where specific user-item interactions have not been previously encountered.

Formally, we define $\mathcal{P}_u$ and $\mathcal{P}_i$ as the generated text-based profiles of user $u$ and item $i$, respectively. These profiles are then encoded into representations $\mathbf{e}_u$ and $\mathbf{e}_i$ using a language model (LM). This process is shown as:
\begin{align}
    \mathbf{e}_u = \text{LM}(\mathcal{P}_u), ~~~ 
    \mathbf{e}_i = \text{LM}(\mathcal{P}_i).\label{eq:1}
\end{align}
The preference score $p_{u,i}$ between user $u$ and item $i$ is calculated as the cosine similarity between their text embeddings $\mathbf{e}_u$ and $\mathbf{e}i$. This is expressed as $p_{u,i} = \cos(\mathbf{e}_u, \mathbf{e}_i)$. We then recommend the top-$k$ unclicked items with highest similarity scores, resulting in a recommendation set.
\begin{align}
    \mathcal{R}_u = \text{top\text{-}k}_{i \in \mathcal{I} \setminus \mathcal{N}_u} cos(\mathbf{e}_u, \mathbf{e}_i).\label{eq:2}
\end{align} \\\vspace{-0.2in}

\noindent \textbf{Text-enhanced Collaborative Filtering}. Collaborative filtering (CF) is a widely used recommendation paradigm. It leverages the collaborative relationships among users and items. This existing CF paradigm can be enhanced by integrating encoded semantic representations. Typically, the value $p_{u, i}$ is calculated based on the interaction data. This is expressed as $p_{u, i} = f(u, i, \mathcal{A})$, where $\mathcal{A}$ represents the interaction data. Text-enhanced collaborative filtering builds upon this foundation. It incorporates textual features $\mathbf{e}$ encoded by language models as supplementary representations. This integration aims to improve the recommendation performance of traditional ID-based frameworks.
\begin{align}
    p_{u, i} = f(u, i, \mathcal{A}, \mathbf{e}_u, \mathbf{e}_i).
\end{align}

\section{Methodology}
\label{sec:solution}

In this section, we first discuss how we gather textual profiles for users and items. Next, we dive into the specifics of \model\ and its training approach. Lastly, we introduce our method for diversifying user profiles. This method improves the model's ability to adapt to various situations.

\subsection{Collaborative User and Item Profiling}\label{sec:profile}

In real-world recommenders, only raw text data may be available. This includes item titles and categories. Privacy concerns limit comprehensive user information. Directly using this textual data can overlook essential collaborative relationships. These relationships are needed for effective user behavior modeling. To address these issues, we propose generating textual profiles using large language models. Examples include GPT and LLaMA series~\cite{RLMRec}. These models incorporate collaborative information.

\subsubsection{\textbf{Item Profiling}}

Given raw item information, we aim to generate a comprehensive item profile $\mathcal{P}_i$. This information includes title $h_i$, categories $c_i$, and description $d_i$ (e.g., book summary). The profile captures both semantic and collaborative aspects. To reflect user-item interactions, we incorporate textual information. This includes user reviews $r_{u, i}$. The item profile generation process is:
\begin{align}
    \mathcal{P}_i = \text{LLM}(\mathcal{M}_i, h_i, c_i, \{r_{u, i}\})\vee\text{LLM}(\mathcal{M}_i, h_i, d_i),\label{eq:item profile}
\end{align}
Here, $\mathcal{M}_i$ depicts the generation instruction. We consider two scenarios. One includes a description in the raw data (right-hand side). The other does not include a description (left-hand side). In the latter case, we utilize collaborative reviews to inform the profile. Using LLMs, we can generate an informative item profile $\mathcal{P}_i$.

\subsubsection{\bf{User Profiling}} 

In practical scenarios, privacy concerns limit generating user profiles from demographic information. Instead, we profile users by leveraging their collaborative relationships through interacted item profiles. This approach captures collaborative signals that reflect user preferences. The user profile generation process is defined as:
\begin{align}
    \mathcal{P}_u = \text{LLM}(\mathcal{M}_u, \{ h_i, \mathcal{P}_i, r_{u,i} \mid i \in \mathcal{N}_u \}).\label{eq:user profile}
\end{align}
Here, $\mathcal{M}_u$ is the instruction for using a LLM to generate the user profile. We sample interacted items $\mathcal{N}_u$ from the user's purchase history. We combine their feedback $r_{u,i}$ with the pre-generated item profiles $\mathcal{P}_i$. This creates the user's text description $\mathcal{P}_u$, capturing their preferences. The incorporation of sampled data ensures that LLMs accurately infer profiles. This includes item metadata and user reviews. The profiles authentically reflect users' interaction preferences. To enhance profile quality, we adopt the principles of Chain of Thought (CoT)\cite{wei2022chain} and Self-Consistency\cite{wang2022self}. We require LLMs to generate explanatory rationales alongside the profiles. It improves inference reliability. It also establishes an interpretable connection between collaborative signals and profile generation.

\subsubsection{\bf{Advantages of Collaborative Profiling}}\label{sec:benefit} Our collaborative profiling framework offers two key advantages for real-world recommendation:
\begin{itemize}[leftmargin=*]
    \item \textbf{Preservation of Collaborative Information}.
    Our approach captures the original textual content. It also captures the semantics of user/item characteristics and their interaction patterns. We encode these profiles into a shared feature space. This uses a recommendation-oriented language model. The embeddings of interacted users and items are aligned. This allows recommenders to better identify relevant matches.\vspace{-0.05in}
    
    \item \textbf{Rapid Adaptation to Dynamic Scenarios}.
    Our profiling enables the recommender system to adapt to evolving user preferences. It also adapts to changing interaction patterns. With robust language models, simple updates to textual user profiles can quickly reflect shifts in interests. They can also reflect shifts in behaviors. This flexibility makes our approach ideal for environments. User interests change over time in environments.
\end{itemize}

\subsection{Profile Embedder with Collaborative LM}
We have generated rich textual profiles for users and items, moving beyond conventional ID-based embeddings. However, directly encoding these textual profiles into latent embeddings for recommendations has two key limitations: 
\begin{itemize}[leftmargin=*]
\item \textbf{Capturing Recommendation-Specific Semantics}. While text embeddings are expressive, they may not optimize for the specific semantics relevant to recommendations. For example, consider two user profiles: (i) "This user is passionate about advanced AI techniques, focusing on deep learning and research." (ii) "With a passion for advanced AI development, this user enjoys science fiction and AI-themed novels." Although both profiles mention AI, their target audiences differ—one caters to AI scientists, while the other targets sci-fi readers. Directly encoding these profiles may overlook recommendation-specific semantics, necessitating refinement to align embeddings with the context of recommendation.\vspace{-0.08in}
\item \textbf{Overlooking High-Order Collaborative Signals}.
While textual profiles provide rich semantic information, relying solely on them may cause us to miss valuable high-order collaborative patterns. These patterns arise from complex user-item interactions~\cite{wang2019neural,xia2023automated}. Such signals include transitive associations and community-level preferences. They offer additional insights that enhance preference learning.
\end{itemize}

To address these limitations, we propose a collaborative language modeling paradigm. It integrates the semantic richness of profiles with valuable collaborative signals from complex interactions.

\subsubsection{\bf Bidirectional Transformer Embedder} 
We use a multi-layer bidirectional Transformer encoder as the backbone for two key benefits: 1) \textbf{Efficient Encoding}: The encoder-only architecture generates effective text representations, allowing faster inference in recommendation. 2) \textbf{Flexible Adaptation}: Leveraging pre-trained Transformer models enables us to optimize the embedder for specific recommendation tasks effectively.

Let's consider a user's profile as a passage of $n$ words: $\mathcal{P} = {w_1, \ldots, w_n}$. We start by adding a special token [CLS] at the beginning of the word sequence. The tokenization layer $\text{Tok}(\cdot)$ then encodes the input sequence into initial embeddings, which serve as the input for the Transformer layers:
\begin{align}
    \{\mathbf{x}_{\texttt{[CLS]}}^{(0)}, \ldots \mathbf{x}_{n}^{(0)}\} = \text{Tok}(\{w_{\texttt{[CLS]}}, \ldots, w_n\}).
\end{align}
Here, $x^{(0)} \in \mathbb{R}^d$ is the embedding from the embedding table for the tokens, with the $(0)$ superscript indicating it is the input to the $(0)$-th layer of the language model. The tokenization process also adds positional embeddings. The language model then encodes a sequence of final embeddings:
\begin{align}
    \{\mathbf{e}_{[CLS]}, \ldots, \mathbf{e}_{n}\} = \text{Enc}(\{\mathbf{x}_{\texttt{[CLS]}}^{(0)}, \ldots \mathbf{x}_{n}^{(0)}\}),
\end{align}
where $\text{Enc}(\cdot)$ refers to the Transformer-based encoder-only LM. The key operation in the encoding process is the self-attention mechanism:
\begin{equation}
\begin{split}
    \text{Attention}(Q, K, V) = \text{softmax}({QK^T}/{\sqrt{d}})V \\
    \textit{w.r.t.}\, Q = XW^Q,\, K = XW^K,\, V = XW^V.
\end{split}
\end{equation}

Here, $X \in \mathbb{R}^{n \times d}$ represents the stack of token embeddings, while $W^{Q/K/V}$ are parameter matrices that map these embeddings into queries, keys, and values. This self-attention mechanism enables each token to aggregate information from all others, ensuring awareness of the entire sequence. We then select the first embedding $\mathbf{e}_{\texttt{[CLS]}}$ as the representative embedding for the profile. This embedding is passed through a multi-layer perceptron to obtain the encoded representation $e$, as shown in Eq.(\ref{eq:1}):
\begin{align}
    \mathbf{e} = \text{MLP}(\mathbf{e}_{\texttt{[CLS]}}) = \text{LM}(\mathcal{P}).
\end{align}
With these encoded text embeddings $\mathbf{e}$ for each user and item, we can predict the score of interaction using cosine similarity and make recommendations as described in Eq.(\ref{eq:2}).

\begin{figure}[t]
    \centering
    \includegraphics[width=1.0\columnwidth]{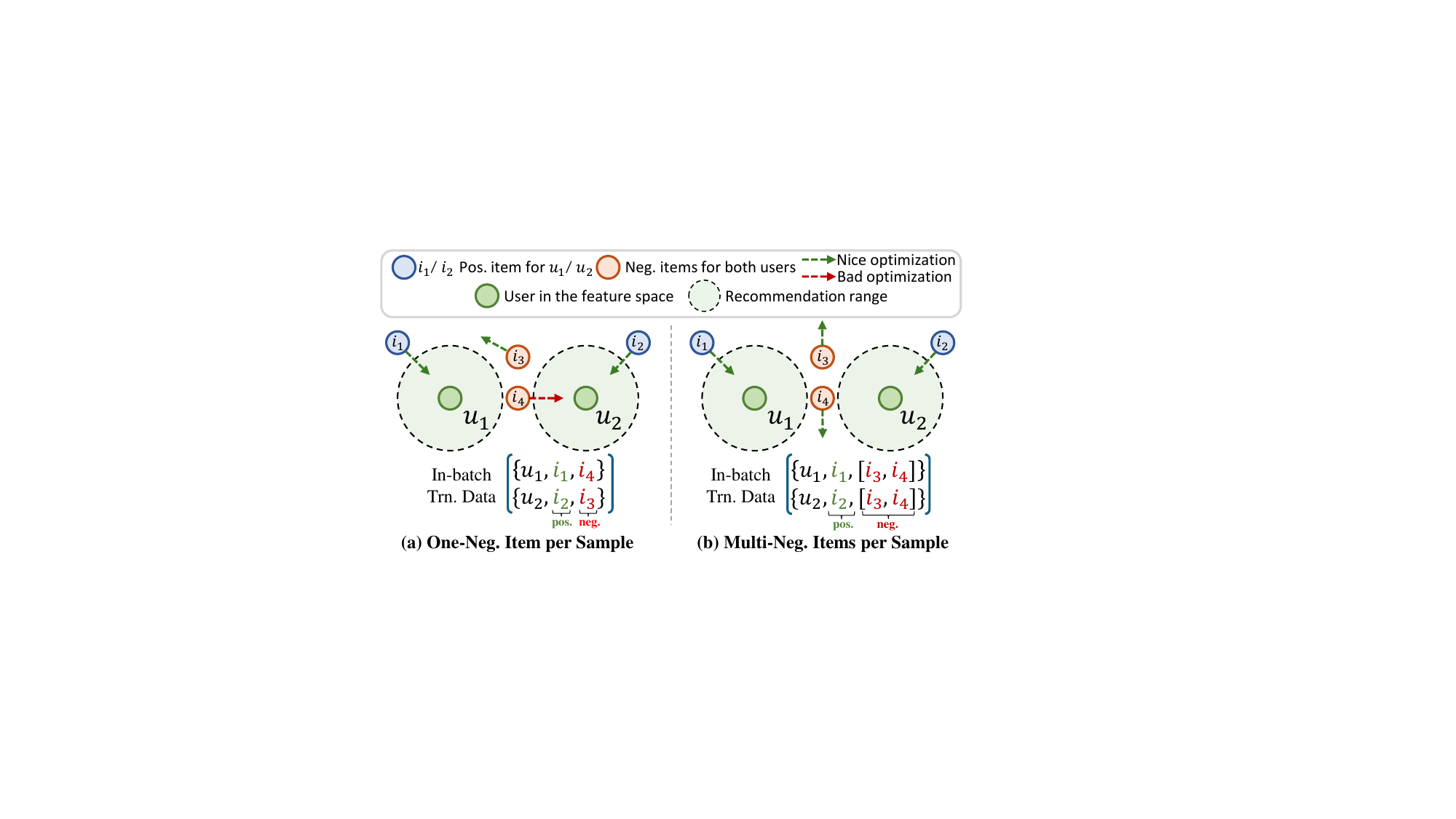}
    \vspace{-0.1in}
    \caption{Contrastive tuning of the collaborative LM enables it to learn rich representations. It aligns the text-based semantic space with global collaborative signals.}
    \vspace{-0.1in}
    \label{fig:intuition}
\end{figure}

\subsubsection{\textbf{Contrastive Collaborative LM Learning}}
We fine-tune the collaborative LM using contrastive learning to effectively capture high-order collaborative signals. Traditional recommenders using Bayesian Personalized Ranking (BPR)~\cite{rendle2012bpr} optimize embeddings with only one negative item per sample. This limits their ability to capture complex global user-item relationships.

The supervised contrastive loss offers a strong alternative to traditional methods. By treating interacted user-item pairs as positives and non-interacted pairs as negatives, it brings related item embeddings closer in the feature space. As shown in Figure~\ref{fig:intuition}, contrastive learning uses a batch of negatives for a comprehensive adjustment of the encoded space, enabling the model to capture high-order collaborative relationships. We evaluate this in Appendix~\ref{apd:objectives} to assess the impact of different learning objectives. The learning objective for tuning the language model is expressed as follows:
\begin{align}
    \mathcal{L}_{\text{con}} = -\sum \log\frac{\exp(s_{u,i^+}/\tau)}{\sum_{j \in \mathcal{N}^-}\exp(s_{u,j}/\tau)},
\end{align}
where $s_{u,j} = \cos(\mathbf{e}_u, \mathbf{e}_j)$ denotes the cosine similarity between user $u$ and item $j$, $i^+$ represents the positive item that user $u$ has interacted with, $\tau$ is a temperature hyperparameter that controls the degree of learning, and $\mathcal{N}^-$ is the set of in-batch negative items.
We also build on prior work~\cite{RecFormer, BLaIR} by incorporating an auxiliary masked language modeling (MLM) loss $\mathcal{L}_{\text{mlm}}$. This technique randomly masks input tokens, training the model to predict them, which stabilizes training and enhances generalization. The final training objective combines the contrastive loss and the MLM loss:
\begin{align}
    \mathcal{L} = \mathcal{L}_{\text{con}} + \lambda \mathcal{L}_{\text{mlm}}.
\end{align}
$\lambda$ is a hyperparameter that balances contrastive learning with masked language modeling loss.

\subsection{Augmentation with Profile Diversification}\label{sec:augmentation}
To enhance generalization to unseen users/items, we propose profile diversification. Single profiles per user or item limit representation diversity and training quality, hurting performance. Our augmentation creates multiple profiles per entity while preserving semantic meaning—capturing personalized preferences for users and varied characteristics.

Inspired by self-instruction mechanisms~\cite{wang2022self,xu2023wizardlm}, LLMs can rephrase user or item profiles while preserving their underlying meaning. This generates multiple semantically similar yet distinctly worded profiles from single inputs. Iterative rephrasing creates diverse augmented profiles, substantially expanding available training data. This technique proves particularly valuable for limited datasets, as LLM-generated profiles improve model generalization and robustness. Through LLM-based diversification, we create diverse profile sets for each user and item.
\begin{align}
    \{ \mathcal{P} \}_{u} &= \{ \mathcal{P}_u;\,\, \mathcal{P}_{u}^1,\, \mathcal{P}_u^2,\, \ldots,\, \mathcal{P}_u^t \}, \\
    \{ \mathcal{P} \}_{i} &= \{ \mathcal{P}_i;\,\, \mathcal{P}_{i}^1,\, \mathcal{P}_i^2,\, \ldots,\, \mathcal{P}_i^t \}.
\end{align}
Here, $\mathcal{P}_{u/i}$ represents the original profile, while $\mathcal{P}_{u/i}^{1-t}$ denotes the LLM-rephrased profiles, with $t$ indicating the number of diversification steps. During training, we randomly select one profile from the user's or item's profile set for each batch data.

\section{Evaluation}
\label{sec:eval}

We evaluate the \model\ in addressing the following research questions (RQs): \textbf{RQ1}: How effectively does the \model\ perform in matching unseen users and items (zero-shot) within text-based recommendation? \textbf{RQ2:} How effectively does \model\ integrate to enhance recommendations in text-based collaborative filtering scenarios? \textbf{RQ3}: How effective is our profile diversification mechanism for augmenting data and improving the recommendation model's performance? \textbf{RQ4}: How well can our proposed text-based \model\ adapt to changes in users' dynamic preferences?

\subsection{Experimental Settings}

\begin{table}[t]
\small
\centering
\caption{Dataset statisics: "Avg. n" is the average interactions per user, and "Inters." stands for interactions. Datasets with \underline{underlines} are from different platforms.}
\vspace{-0.15in}
\label{tab:stat}
\begin{tabular}{lrrrr}
\toprule
\multicolumn{1}{c}{\textbf{Datasets}} & \multicolumn{1}{c}{\textbf{\#Users}} & \multicolumn{1}{c}{\textbf{\#Items}} & \multicolumn{1}{c}{\textbf{\#Inters.}} & \multicolumn{1}{c}{\textbf{Avg. n}} \\ \midrule
\textbf{Train Data} & 124,732 & 67,455 & 802,869 & 6.44 \\
-Arts        & 14,470 & 8,537   & 96,328    & 6.66 \\
-Games       & 17,397 & 8,330   & 120,255   & 6.91 \\
-Movies      & 16,994 & 9,370   & 134,649   & 7.92 \\
-Home        & 22,893 & 13,070  & 131,556   & 5.75 \\
-Electronics & 26,837 & 14,033  & 165,628   & 6.17 \\
-Tools       & 26,141 & 14,155  & 154,453   & 5.91 \\
\midrule
\textbf{Test Data} & 55,877 & 28,988 & 615,210 & 11.01 \\
-Sports             & 21,476 & 12,741  & 132,400 & 6.17  \\
-\underline{Steam}  & 23,310 & 5,237   & 316,190 & 13.56 \\ 
-\underline{Yelp}   & 11,091 & 11,010  & 166,620 & 15.02 \\ 
\bottomrule
\end{tabular}
\vspace{-0.1in}
\end{table}

\begin{table}[t]
\footnotesize
\centering
\caption{Model variants of \model\ differ by parameter size. "HS" stands for hidden size.}
\vspace{-0.1in}
\label{tab:variants}
\begin{tabular}{lcccc}
\toprule
\multicolumn{1}{c}{\textbf{Model}} & \multicolumn{1}{c}{\textbf{Layers}} & \multicolumn{1}{c}{\textbf{HS}} & \multicolumn{1}{c}{\textbf{Heads}} & \multicolumn{1}{c}{\textbf{Params}} \\ 
\midrule
\model-Small   & 6   & 768   & 12 & 82M    \\
\model-Base    & 12  & 768   & 12 & 125M   \\
\model-Large   & 24  & 1024  & 16 & 355M   \\
\bottomrule
\end{tabular}
\vspace{-0.1in}
\end{table}

\begin{table*}[t]
  \centering
  \small
  \caption{Text-based recommendation performance of various LMs across different datasets. The best performance is indicated in \textbf{bold}, while the second-best is highlighted with \underline{underline}. The script $*$ denotes significance (p < 0.05).}
  \vspace{-0.1in}
  \resizebox{\textwidth}{!}{
    \begin{tabular}{c|cccc|cccc|cccc}
      \toprule
            Data & \multicolumn{4}{c|}{Sports} & \multicolumn{4}{c|}{Steam} & \multicolumn{4}{c}{Yelp} \\
        \midrule
            Methods & Recall@10 & Recall@20 & NDCG@10 & NDCG@20 & Recall@10 & Recall@20 & NDCG@10 & NDCG@20 & Recall@10 & Recall@20 & NDCG@10 & NDCG@20\\\midrule
        \multicolumn{13}{c}{\textit{Proprietary Models}} \\
        \midrule
            OpenAIv3-Small   & 0.0324 & 0.0444 & 0.0198 & 0.0230 & 0.0066 & 0.0119 & 0.0049 & 0.0068 & 0.0028 & 0.0056 & 0.0021 & 0.0031\\
            OpenAIv3-Large   & 0.0300 & 0.0436 & 0.0180 & 0.0217 & 0.0070 & 0.0137 & 0.0049 & 0.0073 & 0.0029 & 0.0055 & 0.0023 & 0.0032\\
        \midrule
            \multicolumn{13}{c}{\textit{Base Size Models}} \\
        \midrule
            BERT-Base      & 0.0015 & 0.0032 & 0.0008 & 0.0013 & 0.0015 & 0.0031 & 0.0011 & 0.0017 & 0.0009 & 0.0018 & 0.0006 & 0.0010\\
            RoBERTa-Base   & 0.0121 & 0.0206 & 0.0065 & 0.0087 & 0.0041 & 0.0078 & 0.0031 & 0.0043 & 0.0018 & 0.0034 & 0.0014 & 0.0020\\
            BART-Base      & 0.0088 & 0.0143 & 0.0048 & 0.0063 & 0.0033 & 0.0062 & 0.0024 & 0.0035 & 0.0014 & 0.0027 & 0.0012 & 0.0016\\
            SimCSE-Base    & 0.0240 & 0.0341 & 0.0143 & 0.0170 & 0.0048 & 0.0091 & 0.0035 & 0.0050 & 0.0020 & 0.0039 & 0.0015 & 0.0022\\
            BLaIR-Base     & 0.0251 & 0.0352 & 0.0145 & 0.0173 & 0.0041 & 0.0081 & 0.0030 & 0.0044 & 0.0025 & 0.0046 & 0.0019 & 0.0026\\
            GTR-Base       & 0.0290 & 0.0417 & 0.0172 & 0.0206 & 0.0075 & 0.0136 & 0.0056 & 0.0077 & 0.0025 & 0.0052 & 0.0019 & 0.0029\\
            BGE-Base       & 0.0315 & 0.0437 & 0.0194 & 0.0227 & 0.0094 & 0.0170 & 0.0069 & 0.0095 & 0.0029 & 0.0053 & 0.0022 & 0.0031\\
        \midrule
            \multicolumn{13}{c}{\textit{Large Size Models}} \\
        \midrule
            BERT-Large      & 0.0011 & 0.0021 & 0.0006 & 0.0008 & 0.0019 & 0.0040 & 0.0014 & 0.0022 & 0.0008 & 0.0019 & 0.0007 & 0.0010\\
            RoBERTa-Large   & 0.0039 & 0.0065 & 0.0021 & 0.0028 & 0.0027 & 0.0052 & 0.0021 & 0.0029 & 0.0012 & 0.0023 & 0.0010 & 0.0014\\
            BART-Large      & 0.0114 & 0.0170 & 0.0064 & 0.0080 & 0.0036 & 0.0065 & 0.0026 & 0.0037 & 0.0017 & 0.0033 & 0.0014 & 0.0019\\
            SimCSE-Large    & 0.0232 & 0.0328 & 0.0134 & 0.0160 & 0.0051 & 0.0095 & 0.0037 & 0.0052 & 0.0023 & 0.0044 & 0.0017 & 0.0025\\
            BLaIR-Large     & 0.0227 & 0.0322 & 0.0133 & 0.0159 & 0.0057 & 0.0108 & 0.0041 & 0.0059 & 0.0027 & 0.0047 & 0.0019 & 0.0027\\
            GTR-Large       & 0.0329 & 0.0446 & 0.0198 & 0.0229 & 0.0095 & 0.0168 & 0.0069 & 0.0094 & 0.0021 & 0.0042 & 0.0018 & 0.0025\\
            BGE-Large       & 0.0324 & 0.0449 & 0.0196 & 0.0230 & 0.0089 & 0.0153 & 0.0066 & 0.0088 & 0.0025 & 0.0052 & 0.0020 & 0.0029\\
        \midrule
            \multicolumn{13}{c}{\textbf{\textit{\model\ Series}}} \\
        \midrule
            \model-Small & 0.0186 & 0.0286 & 0.0108 & 0.0135 & 0.0097 & 0.0174 & 0.0070 & 0.0097 & 0.0022 & 0.0046 & 0.0017 & 0.0026\\
            \model-Base  & \underline{0.0360} & \underline{0.0518} & \underline{0.0210} & \underline{0.0253} & \underline{0.0114} & \underline{0.0203} & \underline{0.0081} & \underline{0.0112} & \underline{0.0034} & \underline{0.0063} & \underline{0.0026} & \underline{0.0037}\\
            \model-Large & \textbf{0.0396}* & \textbf{0.0557}* & \textbf{0.0236}* & \textbf{0.0279}* & \textbf{0.0129}* & \textbf{0.0225}* & \textbf{0.0093}* & \textbf{0.0127}* & \textbf{0.0034}* & \textbf{0.0065}* & \textbf{0.0026}* & \textbf{0.0037}*\\
        \midrule
            \textbf{\textit{Improve}} & $\uparrow$ 20.36\% & $\uparrow$ 24.05\% & $\uparrow$ 19.19\% & $\uparrow$ 21.30\% & $\uparrow$ 35.79\% & $\uparrow$ 32.35\% & $\uparrow$ 39.13\% & $\uparrow$ 33.68\% & $\uparrow$ 17.24\% & $\uparrow$ 16.07\% & $\uparrow$ 13.04\% & $\uparrow$ 15.63\%\\
      \bottomrule
      \end{tabular}
    }
    \vspace{-0.15in}
  \label{tab:overall comparison}%
\end{table*}%

\subsubsection{\textbf{Datasets}}\label{sec:data} To assess our proposed model's capability in encoding user/item textual profiles into embeddings for recommendation, we curated diverse datasets across various domains and platforms. A portion was used for training, while the remainder served as test sets for zero-shot evaluation. The dataset statistics are shown in Table~\ref{tab:stat}. Due to the page limit, we place the detail of data resources are described in Appendix~\ref{adp:dataset}

\subsubsection{\textbf{Evaluation Protocols}} We employ two ranking-based evaluation metrics, Recall@$N$ and NDCG@$N$, to assess performance in both text-based recommendation and collaborative filtering scenarios. Specifically, we compute these metrics for $N$ values of 10 and 20~\cite{he2017neural, he2020lightgcn}. The evaluation is conducted using the all-rank protocol~\cite{he2020lightgcn} with the predicted preference scores.
In the context of text-based recommendation, particularly for datasets containing multiple LLM-diversified profiles (as discussed in Section~\ref{sec:augmentation}), we calculate the metrics separately $t$ times based on different profile pairs $(\mathcal{P}_u^{1, \ldots, t}, \mathcal{P}_i^{1, \ldots, t})$. Subsequently, we compute the mean value for each metric to obtain a comprehensive assessment. For the training datasets, we utilize the validation split for evaluation, while for the test datasets, we employ the test split.

\subsection{Performance Comparision for Text-based Recommendation (RQ1)}
We evaluate the performance of various language models (LMs) for zero-shot text-based recommendation on the unseen Sports, Steam, and Yelp datasets. This approach directly leverages the encoded embeddings derived from user/item profiles to make recommendations, without any additional training on the target datasets.

\subsubsection{\textbf{Baseline Methods and Settings}} We have included the following state-of-the-art language models as text embedders for comparative evaluation: (i) General Language Models: \textbf{BERT}~\cite{BERT}, \textbf{RoBERTa}~\cite{RoBERTa} and \textbf{BART}~\cite{BART}; (ii) Language Models for Dense Retrieval: \textbf{SimCSE}~\cite{SimCSE}, \textbf{GTR}~\cite{GTR} and \textbf{BGE}~\cite{BGE}; (iii) Pre-trained Language Models for Recommendation: \textbf{BLaIR}~\cite{BLaIR}. We also compare with SoTA text embedding models provided by \textbf{OpenAI}. Details of the baseline models are described in Appendix~\ref{apd:detail}.

\subsubsection{\textbf{Result Analysis}} The overall comparison of different models is presented in Table~\ref{tab:overall comparison}. This evaluation reveals several observations, which are outlined below:
\begin{itemize}[leftmargin=*]

    \item \textbf{Superiority across Diverse Datasets}. Our evaluation consistently shows that the \model\ outperforms all other models across the datasets spanning different platforms. This provides strong evidence for the effectiveness of the \model. We attribute these improvements to two key factors: i) By injecting collaborative signals into the LMs, we effectively optimized our \model\ using supervised contrastive learning within the recommendation context. This approach allows the model to inherently encode user and item text embeddings that are well-suited for recommendation tasks. ii) By integrating a diverse array of datasets across multiple categories and utilizing data augmentation techniques to enrich the text descriptions for training, our \model\ exhibits impressive generalization capabilities, enabling it to effectively handle unseen data. \vspace{-0.12in}

    \item \textbf{Scaling Law Investigation of \model\ Model}. Our experiments reveal that as the size of the \model\ increases, its performance consistently improves across all the datasets. This observation reflects a scaling law, where the model's performance growth is directly correlated with its size. Furthermore, this finding effectively reinforces the validity of text-based recommendation systems. It also validates our approach to training the LMs, which enables it to learn collaborative signals from a new perspective.
    
\end{itemize}

\subsection{Performance of Text-enhanced CF (RQ2)}~\label{sec:text-enhanced CF} \vspace{-0.15in}

\begin{table}[t]
\small
\centering
\caption{Recommendation performance in text-enhanced CF. The experiment was conducted on the Steam dataset with 5-runs to obtain the mean results.}
\vspace{-0.15in}
\label{tab:enhancedcf}
\begin{tabular}{lcc|cc}
\toprule
\multirow{2}{*}{Metric}
& \multicolumn{2}{c|}{Recall} & \multicolumn{2}{c}{NDCG} \\ 
& @10 & @20 & @10 & @20 \\
\midrule
\multicolumn{5}{l}{\textbf{ID-based Methods}} \\
GCCF      & 0.0826 & 0.1314 & 0.0665 & 0.0830\\
LightGCN  & 0.0851 & 0.1349 & 0.0686 & 0.0854\\
\midrule
\multicolumn{5}{l}{\textbf{Text-enhanced GCCF}} \\
BERT    & 0.0822 & 0.1313 & 0.0663 & 0.0829 \\
RoBERTa & 0.0848 & 0.1351 & 0.0684 & 0.0854 \\
BART    & 0.0874 & 0.1383 & 0.0701 & 0.0874 \\
SimCSE  & 0.0877 & 0.1395 & 0.0706 & 0.0881 \\
BLaIR   & 0.0880 & 0.1392 & 0.0708 & 0.0882 \\
GTR     & 0.0873 & 0.1387 & 0.0706 & 0.0880 \\
BGE     & 0.0875 & 0.1393 & 0.0705 & 0.0881 \\
\textbf{EasyRec} & \textbf{0.0881} & \textbf{0.1402} & \textbf{0.0712} & \textbf{0.0888} \\
\midrule
\multicolumn{5}{l}{\textbf{Text-enhanced LightGCN}} \\
BERT    & 0.0849 & 0.1347 & 0.0684 & 0.0852 \\
RoBERTa & 0.0867 & 0.1374 & 0.0699 & 0.0870 \\
BART    & 0.0887 & 0.1407 & 0.0715 & 0.0891 \\
SimCSE  & 0.0891 & 0.1417 & 0.0719 & 0.0898 \\
BLaIR   & 0.0897 & 0.1418 & 0.0724 & 0.0901 \\
GTR     & 0.0894 & 0.1417 & 0.0719 & 0.0896 \\
BGE     & 0.0891 & 0.1407 & 0.0718 & 0.0893 \\
\textbf{EasyRec} & \textbf{0.0908} & \textbf{0.1430} & \textbf{0.0732} & \textbf{0.0908} \\
\bottomrule
\end{tabular}
\vspace{-0.15in}
\end{table}

In addition to our investigation of zero-shot recommendation, we explore the potential of \model\ as an enhancement when integrated with CF models. To assess the effectiveness of various LMs in CF, we employ two widely used ID-based methods as backbone models: \textbf{GCCF}~\cite{chen2020revisiting} and \textbf{LightGCN}~\cite{he2020lightgcn}, which were chosen for their proven effectiveness and efficiency. Furthermore, we utilize the advanced model-agnostic text-enhanced framework RLMRec~\cite{RLMRec} with contrastive alignment to conduct our investigation. We compare the large versions of both \model\ and other open-source LMs. The findings from the results in Table~\ref{tab:enhancedcf} are: \vspace{-0.15in}

\begin{itemize}[leftmargin=*]
    \item Compared to backbones, the integration of the text-enhanced framework generally improves the performance for both GCCF and LightGCN. This observation highlights the significance of incorporating text modality (\ie, user/item profiles) into the recommendation paradigm. \vspace{-0.15in}
    \item Among the various LMs, \model\ consistently achieves the highest performance in the text-enhanced recommenders. This outcome not only illustrates the efficacy of \model\ for recommendation, but also emphasizes the advantages of incorporating collaborative information into LMs.
\end{itemize}

\subsection{Efficacy of Profile Diversification (RQ3)}

\begin{figure}[t]
    \centering
    \subfigure[Performance on Steam data]{
        \includegraphics[width=1.0\columnwidth]{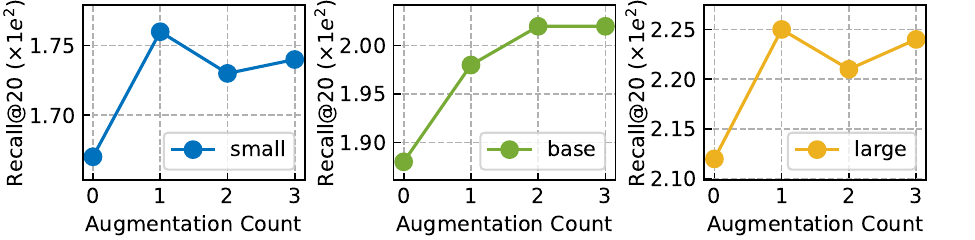}\ 
        \vspace{-0.15in}
    }
    \subfigure[Performance on Yelp data]{
        \includegraphics[width=1.0\columnwidth]{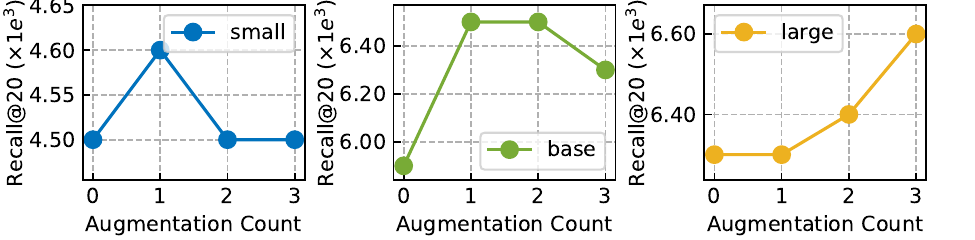}
    }
    \vspace{-0.15in} 
    \caption{Performance \textit{w.r.t.} data size. "Augmentation Count" indicates the number $t$ of diversified profiles.}
    \vspace{-0.2in}
    \label{fig:scale data}
\end{figure}

In this section, we examine the impact of diversifying user and item profiles with large language models (LLMs) on model performance. As mentioned in Section~\ref{sec:augmentation}, we perform LLM-based diversification three times on the original generated profiles. This process continuously increases the number of profiles in the training set. To investigate whether data augmentation positively affects model performance, we conduct experiments with three variants of the \model\ under different numbers of diversified profiles. The results are shown in Figure~\ref{fig:scale data}, leading to the following key observations:\vspace{-0.1in}

\begin{itemize}[leftmargin=*]
    \item \textbf{Effectiveness of Profile Diversification}. The increase in the number of diversified profiles enhances model performance, particularly for larger models. This finding underscores the effectiveness of our augmentation approach using LLMs for profile diversification, and emphasizes the significance of increasing training corpus for improved recommendation outcomes.\vspace{-0.1in}
    
    \item \textbf{Scaling Relationship}: The scaling experiments on both model size and data size reveal a crucial relationship that influences model performance. This demonstrates that our approach of training the LM with collaborative signals follows a scaling law, indicating that model performance benefits from both increased capacity and data volume. Such findings are vital as they provide insights into how model capacity and data availability interact, guiding future research and development.\vspace{-0.1in}
\end{itemize}

Moreover, we find that profile diversification can improve recommendation diversity. We conduct experiments and discussions in Appendix~\ref{apd:diversity}.

\subsection{Fast Adaptation Case Study (RQ4)}\label{sec:case_study}

\begin{figure}[t]
    \centering
    \includegraphics[width=1.0\columnwidth]{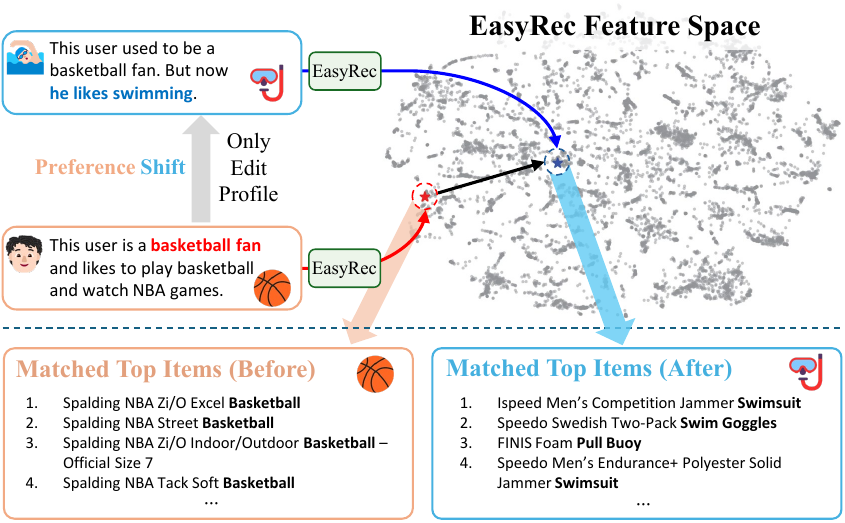}
    \vspace{-0.1in}
    \caption{Case study on handling user preference shift.}
    \vspace{-0.1in}
    \label{fig:case study}
\end{figure}

As mentioned in Section~\ref{sec:benefit}, a key advantage of \model\ is its ability to empower recommender systems to efficiently adapt to shifts in user preferences and behavior dynamics over time. To evaluate this capability, we create two user profiles reflecting shifted preferences on the \textit{Amazon-Sport} dataset and examine the recommended items from the \model. As shown in Figure~\ref{fig:case study}, the original user profile indicates that the user enjoys playing basketball. However, the user's preference later transitions to a preference for swimming.

We visualize all the encoded embeddings using t-SNE~\cite{van2008visualizing}, as illustrated in Figure~\ref{fig:case study}, which reveals a significant shift in the user embedding within the feature space. Correspondingly, recommended items transition from basketball-related products to swimming gear, reflecting the user's changing preferences. Notably, this adjustment is accomplished solely by modifying the user's profile, without further training of the model. This underscores the efficiency and flexibility in adapting to evolving user preferences.
\section{Related Work}

\noindent \textbf{LM-Powered Recommender Systems}. 
Recent recommendation methods increasingly incorporate text modalities~\cite{yuan2023go, wu2024could, zhang2023collm}, using LM-derived semantic embeddings to boost CTR prediction and sequence recommendation~\cite{KAR, geng2024breaking, UniSRec, AlphaRec}. Some approaches use text-based agents to boost performance~\cite{zhang2024generative, zhang2024agentcf}, with notable examples like ZESRec~\cite{ZESRec} and MoRec~\cite{yuan2023go} leveraging text embeddings for inductive performance. RLMRec applies information-theoretic text augmentation to ID-based recommenders~\cite{RLMRec}. 
However, most systems depend on general-purpose encoders (\eg, BERT) rather than recommendation-tailored LMs. Inspired by BLaIR's use of feedback and metadata~\cite{BLaIR}, we train a recommendation-specific LM on user profiles and collaborative signals to improve zero-shot and text-augmented performance.

\noindent \textbf{Cross-Domain Recommendation}. Cross-domain recommendation enhances recommendations in one domain by leveraging data from another domain to combat data sparsity and improve personalization~\cite{zang2022survey}. Techniques like graph collaborative filtering~\cite{liu2020cross} use Graph Neural Networks (GNNs) to aggregate common and domain-specific user features. Recent studies have integrated self-supervised learning, such as C2DSR~\cite{cao2022contrastive} with contrastive learning for both single and cross-domain representations, and CCDR~\cite{xie2022contrastive} with intra- and inter-domain contrastive learning. SITN~\cite{sun2023self} utilizes self-attention to represent user sequences from source and target domains, followed by contrastive learning.
However, current cross-domain approaches often rely on correlations between source and target data, limiting their application to zero-shot tasks. In contrast, our proposed \model\ employs text-based zero-shot learning, removing these constraints and enabling effective transfer across different domains.

\noindent \textbf{Graph Collaborative Filtering for Recommendation}.
Graph Neural Networks (GNNs) are effective for recommendation by modeling user–item interactions and high-order dependencies~\cite{gao2023survey}. Representative models include PinSage~\cite{ying2018graph}, NGCF~\cite{wang2019neural} and LightGCN~\cite{he2020lightgcn}, which learn node representations via neighborhood aggregation. To address data sparsity, recent studies combine self-supervised learning with collaborative filtering using graph augmentations; notable examples are SGL~\cite{wu2021self}, SimGCL~\cite{yu2022graph} and HCCF~\cite{xia2022hypergraph}. These works demonstrate the promise of self-supervised graph learning for recommender systems.
\section{Conclusion}
\label{sec:conclusoin}

This paper presents \model\, which integrates LMs to enhance recommendation. Our approach is simple yet effective, excelling in scenarios like text-based zero-shot recommendations and text-enhanced CF. Central to \model's success is its combination of collaborative LM tuning and contrastive learning, which captures nuanced semantics and high-order collaborative signals, leading to improved recommendations. Extensive experiments demonstrate \model's superiority over existing models, showcasing its adaptability to changing user preferences and real-world applications.
\section*{Acknowledgments}
This work is supported by the National Natural Science Foundation of China under Grants 624B2122.

\clearpage

\section*{Limitation}
\label{sec:limit}

While EasyRec shows promising advancements in generalization for zero-shot recommendation, it faces challenges related to data modality diversity. Currently, our approach relies primarily on textual data, whereas items can encompass a richer variety of modalities. The absence of visual inputs, such as images and videos, limits the contextual information we can leverage. These visual elements have the potential to convey aesthetic preferences, cultural trends, and emotional responses, capturing nuances that textual data might overlook. They can also reveal visual patterns linked to user behavior and preferences, including color schemes, styles, and settings, which are essential for effective personalization. Acknowledging these aspects, future work could explore the integration of multimodal data processing techniques, potentially enhancing predictive accuracy and improving the system's ability to generalize recommendations.

\bibliography{refs}
\appendix
\begin{appendices}

\section{Discussion on \model}\label{apd:discussion}

\subsection{Motivation and Technical Positioning}
The motivation behind our proposed \model\ stems from the limitations of existing recommendation algorithms that primarily rely on ID-based representations. While ID-based representations can enhance algorithm performance, they fundamentally restrict the generalization capabilities of recommendation systems. Specifically, models trained in this manner cannot be directly transferred to new datasets in zero-shot recommendation scenarios, where there is no overlap between users and items. Recent studies~\cite{RLMRec, AlphaRec} have shown a growing interest in incorporating language representations (i.e., text embeddings) to replace or complement ID-based representations. This approach aims to improve the generalization and overall performance of these algorithms. However, the predominant reliance on general-purpose language models as text encoders has created a gap in the development of language models specifically tailored for recommendation scenarios, which could yield higher-quality user and item embeddings. To address this gap, we propose utilizing user and item profiles as fundamental textual representations. By training a language model specifically designed for recommendation algorithms, based on collaborative filtering signals and profile diversification, we can generate embeddings that are more suitable for recommendation contexts. This approach promises to deliver improved performance compared to general LMs.

\subsection{Relationship between \model\ and LLM-CF Methods}
The proposed \model\ constitutes a specialized class of language models meticulously trained for recommendation scenarios. Functioning as a textual encoder, its primary objective lies in transforming descriptive textual profiles of users and items into discriminative feature representations compatible with collaborative filtering architectures. The training process employs collaborative filtering signals as self-supervised signals to guide model optimization. The motivation of \model\ stands distinct from approaches that combine LLMs with CF models as end-to-end frameworks~\cite{RLMRec,yang2024darec}. By contrast, \model\ adopts a foundational modeling perspective, aiming to enhance existing text-enhanced algorithms through improved embeddings.

Recently, LLM-CF integration methods predominantly utilize text embeddings as auxiliary features. As empirically validated in our experiments on Text-enhanced Collaborative Filtering (Section~\ref{sec:text-enhanced CF}), \model\ shows superior performance compared to open-source text embedders.

\subsection{Can Text-Only Frameworks Outperform ID-Based Recommendation?}
ID-based recommendation systems have long served as the predominant paradigm in both industrial deployments and academic research. However, their inherent limitations in cross-domain generalization have recently propelled text-only recommendation approaches -- the core focus of \model\ -- to the forefront of research attention. While text-based methods demonstrate enhanced generalization capabilities, the performance hierarchy between text-based and ID-based approaches remains empirically underexplored.

To investigate this research question, we integrate \model\ with the state-of-the-art text-based framework AlphaRec~\cite{AlphaRec}, conducting comparative analyses against the ID-based LightGCN backbone. To ensure robustness, we perform multiple runs and report averaged results. The experimental results detailed in Table~\ref{tab:text-only comparison} indicate that the text-only framework provides superior performance compared to ID-based methods.

\begin{table}[t]
\footnotesize
\centering
\caption{Performance comparison (Steam dataset) between text-based and ID-based recommenders}
\label{tab:text-only comparison}
\vspace{-0.05in}
\resizebox{0.45\textwidth}{!}{
\begin{tabular}{lcccc}
\toprule
\textbf{Model} & \textbf{R@10} & \textbf{R@20} & \textbf{N@10} & \textbf{N@20} \\
\midrule
LightGCN (ID-based) & 0.0851 & 0.1349 & 0.0686 & 0.0854 \\
\textbf{AlphaRec + \model} & \textbf{0.0863} & \textbf{0.1358} & \textbf{0.0689} & \textbf{0.0857} \\
\bottomrule
\end{tabular}%
}
\vspace{-0.2in}
\end{table}

\begin{figure*}[t]
\centering
\small
\begin{tcolorbox}[title=\texttt{User Profile Diversification}]
\textbf{Instruction}

\texttt{You will assist me in revising a user's profile while maintaining its original meaning. I will present you with the user's initial profile.\\}

\texttt{Instructions:}

\texttt{USER PROFILE: The original user profile.\\}

\texttt{Requirements:}

\texttt{1. Please provide the revised profile directly begin with "REVISED PROFILE: ".}

\texttt{2. The rephrased profile should minimize duplication with the original text while preserving its intended meaning.}

\texttt{3. The revised profile should exhibit varied sentence structures while faithfully conveying the original profile's essence.}

\tikz \draw[dashed] (0,0) -- (\linewidth,0);

\textbf{Input Prompt}

\texttt{USER PROFILE: This user is likely to enjoy items related to baking, entertaining, and colorful table settings. They appreciate convenience, efficiency, and practicality in kitchen appliances.}

\tikz \draw[dashed] (0,0) -- (\linewidth,0);

\textbf{Response}

\texttt{REVISED PROFILE: An individual who finds pleasure in baking, hosting gatherings, and vibrant table arrangements. They value kitchen devices that offer convenience, efficiency, and practicality.}
\end{tcolorbox}
\caption{An example of large language models-based user profile diversification.}\label{fig:user profile diversification}
\end{figure*}

\begin{figure*}[t]
\centering
\small
\begin{tcolorbox}[title=\texttt{Item Profile Diversification}]
\textbf{Instruction}

\texttt{You will assist me in revising a item's profile while maintaining its original meaning. I will present you with the item's initial profile.\\}

\texttt{Instructions:}

\texttt{ITEM PROFILE: The original item profile.\\}

\texttt{Requirements:}

\texttt{1. Please provide the revised profile directly begin with "REVISED PROFILE: ".}

\texttt{2. The rephrased profile should minimize duplication with the original text while preserving its intended meaning.}

\texttt{3. The revised profile should exhibit varied sentence structures while faithfully conveying the original profile's essence.}

\tikz \draw[dashed] (0,0) -- (\linewidth,0);

\textbf{Input Prompt}

\texttt{ITEM PROFILE: The Innovee Lemon Squeezer is a high-quality stainless steel manual citrus press that comes with a lemon recipes ebook. Ideal for those who enjoy fresh lemon juice and recipes.}

\tikz \draw[dashed] (0,0) -- (\linewidth,0);

\textbf{Response}

\texttt{REVISED PROFILE: The Innovee Lemon Squeezer is a stainless steel manual citrus press that includes a lemon recipes ebook, perfect for individuals who appreciate the taste of freshly squeezed lemon juice and love trying out new recipes.}
\end{tcolorbox}
\caption{An example of large language models-based item profile diversification.}\label{fig:item profile diversification}
\end{figure*}

\section{\textbf{Implementation and Training DetaILD}}\label{apd:implementation} We implemented our \model\ and conducted all experiments using PyTorch~\cite{Pytorch}. For the transformer-based encoder backbone, we adopted the architecture of RoBERTa~\cite{RoBERTa} and utilized its pre-trained parameters as initialization. We trained three versions of \model\ with varying parameter sizes (\textbf{small}, \textbf{base}, and \textbf{large}), as detailed in Table~\ref{tab:variants}. 
For the loss function, we set the hyperparameters $\tau$ to 0.05 and $\lambda$ to 0.1. The token masking ratio for masked language modeling is 0.15, and the learning rate is set to $5 \times 10^{-5}$. We train the model for 25 epochs. For profile augmentation, we set the diversification time $t$ for LLM-based methods to 3. During training, we evaluate the model every 1000 steps and use the validation interactions from each training dataset to select the optimal model parameters, employing the Recall@20 metric. Detailed implementation of our model is provided in the code.

\section{Comparison with Zero-shot Frameworks} The exploration of zero-shot recommendation has garnered significant attention, particularly in sequential recommendation scenarios~\cite{ZESRec}. In our work, we propose a novel zero-shot research line for collaborative filtering (CF) based on semantic profile matching. To demonstrate its effectiveness, we compare it with another fundamental zero-shot framework, LLMRank~\cite{hou2024large}, which leverages LLMs for item ranking. Specifically, we implement LLMRank with \texttt{GPT-4o-mini} to rank candidate items based on historical interactions and item titles, while our method employs user/item profile embeddings for feature-based ranking. Following the same experimental protocol, for each user in the test set we construct a candidate set containing 19 negative items and 1 ground-truth item. As shown in Table~\ref{tab:zero_shot_framework}, our approach demonstrates better recall performance while achieving higher efficiency.

\begin{table}[t]
\footnotesize
\centering
\caption{Performance comparison with other zero-shot frameworks on sports dataset.}
\label{tab:zero_shot_framework}
\vspace{-0.05in}
\resizebox{0.45\textwidth}{!}{%
\begin{tabular}{lccc}
\toprule
\textbf{Model} & \textbf{R@5} & \textbf{R@10} & \textbf{Inference Time} \\
\midrule
LLMRank & 0.6155 & 0.7822 & $\sim$1 hours \\
\textbf{EasyRec} & \textbf{0.6332} & \textbf{0.8078} & \textbf{3-4 minutes} \\
\bottomrule
\end{tabular}%
}
\end{table}

\section{Profile Generation with Different Large Language Models} To explore whether the trained \model\ can generalize to different LLMs that under different tokenizers, we regenerate alternative user/item profiles using DeepSeek-V3~\cite{liu2024deepseek} on the Sports dataset and performed a performance validation based on EasyRec-Large. The results on Sports dataset are shown in Table~\ref{tab:llm_profiles}, from which we can observed that the profiles generated based on GPT-3.5 and those generated by DeepSeek have a minimal impact on performance. This is because the final profiles are presented in text form. During the training of EasyRec, we exposed it to a large diversity of recommendation-related texts and utilized multiple datasets for joint optimization, effectively avoiding overfitting on specific profile patterns.

\begin{table}[t]
\footnotesize
\centering
\caption{Performance comparison with various LLM-generated profiles on sports dataset.}
\label{tab:llm_profiles}
\vspace{-0.05in}
\resizebox{0.45\textwidth}{!}{%
\begin{tabular}{lcccc}
\toprule
\textbf{Profile Generator} & \textbf{R@10} & \textbf{R@20} & \textbf{N@10} & \textbf{N@20} \\
\midrule
GPT-3.5 & 0.0423 & 0.0586 & 0.0250 & 0.0294 \\
DeepSeek-V3 & 0.0417 & 0.0555 & 0.0253 & 0.0290 \\
\bottomrule
\end{tabular}%
}
\end{table}

\section{Impact of Training Objectives}\label{apd:objectives}
To evaluate the impact of different training objectives on the LM's learning, we also implemented \model-Large training with BPR loss (i.e., one negative item per training sample) for comparison with the contrastive learning results. As shown in Table~\ref{tab:compareloss}, the model performance generally outperforms that trained with BPR loss, demonstrating the effectiveness of using contrastive learning to incorporate collaborative information into the LMs.

\begin{table}[t]
\small
\centering
\caption{Comparison of \model\ with different learning objectives (where "Contrast" stands for contrastive).}
\vspace{-0.1in}
\label{tab:compareloss}
\begin{tabular}{lcccc}
\toprule
\multirow{2}{*}{\textbf{Objective}}
& \multicolumn{2}{c}{\textbf{Sports}} & \multicolumn{2}{c}{\textbf{Yelp}} \\ 
& \textbf{R@10} & \textbf{N@10} & \textbf{R@10} & \textbf{N@10} \\
\midrule
BPR Loss       & 0.0381 & 0.0226 & 0.0028 & 0.0021 \\
Contrast Loss  & \textbf{0.0396} & \textbf{0.0236} & \textbf{0.0034} & \textbf{0.0026} \\
\bottomrule
\vspace{-0.2in}
\end{tabular}
\end{table}

\section{Datasets and User/Item Profiles}
In this section, we provide detailed information on the processes for profile generation and diversification, including instructions, examples, and associated costs.
\subsection{\textbf{DetaILD of Dataset}}\label{adp:dataset}
We utilize Amazon review data~\cite{ni2019justifying} across six categories to form the training data: \textit{Arts, Crafts and Sewing} (\textbf{Arts}), \textit{Movies and TV} (\textbf{Movies}), \textit{Video Games} (\textbf{Games}), \textit{Home and Kitchen} (\textbf{Home}), \textit{Electronics} (\textbf{Electronics}), and \textit{Tools and Home Improvement} (\textbf{Tools}). For the test datasets, we use one domain, \textit{Sports and Outdoors} (\textbf{Sports}), from the Amazon review data, along with two cross-platform datasets: \textbf{Steam} and \textbf{Yelp}, for comprehensive evaluation. For the datasets from the Amazon platform, we first filter the data to include only those with a rating score $\geq$ 3 and apply a 10-core filtering to densify the dataset. Subsequently, for each category, we split the interactions into training, validation, and test splits in a ratio of 8:1:1. In contrast, for the Steam and Yelp datasets, we directly use the data processed in RLMRec~\cite{RLMRec}, which maintains a split ratio of 3:1:1. 

\subsection{\textbf{DetaILD of Profile Generation}} \label{apd:profile generation}
After the data processing described in Section~\ref{sec:data}, each dataset contains a split of training interactions. We use these interactions to generate user and item profiles according to the paradigm outlined in Section~\ref{sec:profile}, as this requires user-item interaction information. For the Steam and Yelp datasets, we directly utilize the profiles generated by RLMRec~\cite{RLMRec}, which follow the protocol that leverages review information for user and item profile generation. It is important to note that the profiles for each dataset are generated exclusively based on the training interactions. This approach ensures that validation and test interactions are reserved for evaluation purposes, preventing data leakage and allowing for a more accurate assessment of the model's generalization performance on unseen recommendation data.

The profiling process adopts an item-to-user paradigm, where we first generate item profiles in parallel using multi-thread processing, followed by the parallel generation of user profiles that incorporate collaborative information. The large language model employed is \texttt{GPT-3.5-Turbo} from OpenAI.
Specifically, in this work, we process our own datasets from Amazon review data~\cite{ni2019justifying}, which include the categories Arts, Movies, Games, Home, Electronics, Tools, and Sports. For \textbf{item profile generation}, each item $i$ includes a title $h_i$ and an original description $d_i$. We leverage these two pieces of information to generate the profile, as described on the right-hand side of Eq.~\ref{eq:item profile}. Next, for \textbf{user profile generation}, we uniformly sample a number of interacted items from each user's behavior history as collaborative information. We then arrange the input prompt for the large language model according to Eq.~\ref{eq:user profile}. All prompts and instructions for user and item profile generation are provided in the code.

\subsection{\textbf{DetaILD of Profile Diversification}}\label{apd:profile generation}
As described in Section~\ref{sec:augmentation}, we also conduct profile diversification using large language models (LLMs) to enhance the diversity of the training and test datasets, thereby improving and better evaluating the model's generalization ability across different user and item profiles. For each user or item, we perform \(t\) iterations of diversification starting from the initially generated profile. This means that we obtain the first diversified profile based on the original profile and then use this diversified profile for further diversification with the LLMs.

For reference, examples of user and item profile diversification are provided in Figure~\ref{fig:user profile diversification} and Figure~\ref{fig:item profile diversification}, respectively. As illustrated in the case of user profile diversification, the profiles for the same user differ at the word level while still representing the same preferences. Such diversification enhances the quality and diversity of textual data while increasing the overall dataset size.

\begin{table}[t]
\small
\centering
\caption{Costs for profile generation and diversification, including three iterations of diversification.}
\vspace{-0.1in}
\label{tab:cost}
\begin{tabular}{lcccc}
\toprule
\textbf{Operation} & \textbf{\#Data} & \textbf{\#Tokens}  & \textbf{\#Cost (\$)}\\
\midrule
Generation         & 7 & 189M & $\sim$114\\
Diversification    & 9 & 132M & $\sim$97\\
\bottomrule
\vspace{-0.2in}
\end{tabular}
\end{table}

\subsection{\textbf{Cost of Generation and Diversification}} We summarize the total number of tokens and the costs for utilizing LLMs for profile generation and diversification in Table~\ref{tab:cost}. The profiles for the Steam and Yelp datasets have already been generated in previous work~\cite{RLMRec}; therefore, the number of profiled datasets and diversified datasets differs. The total number of tokens required to process both profile generation and diversification is approximately 322 million, costing approximately 200 dollars with the \texttt{GPT-3.5-Turbo} API, making it a cost-effective choice.

\subsection{Improved Recommendation Diversity with Profile Diversification\label{apd:diversity}} To further demonstrate the effecitveness of profile diversification, We also conduct experiments to explore whether diversification can enhance the diversity of recommendation results. We calculated Intra-List Distance (ILD)~\cite{zhang2008avoiding} for \model-Large with and without profile diversification on the Steam and Yelp datasets. As shown in Table~\ref{tab:diversity}, our findings indicate that profile diversification effectively increases the diversity of the model's recommendation results.

\begin{table}[t]
\footnotesize
\centering
\caption{Diversity improvement with profile diversification (Higher ILD indicates better diversity)}
\label{tab:diversity}
\vspace{-0.05in}
\resizebox{0.45\textwidth}{!}{%
\begin{tabular}{lcccccc}
\toprule
\multirow{2}{*}{\textbf{Model}} & \multicolumn{3}{c}{\textbf{Steam}} & \multicolumn{3}{c}{\textbf{Yelp}} \\
\cmidrule(lr){2-4} \cmidrule(lr){5-7}
 & ILD@5 & ILD@10 & ILD@20 & ILD@5 & ILD@10 & ILD@20 \\
\midrule
w/o Diversification & 0.7128 & 0.7336 & 0.7573 & 0.5954 & 0.6090 & 0.6248 \\
w/ Diversification & 0.7268 & 0.7487 & 0.7727 & 0.6012 & 0.6136 & 0.6285 \\
\bottomrule
\end{tabular}%
}
\vspace{-0.2in}
\end{table}

\section{Details of Text-based Recommendation}\label{apd:detail}

\begin{table}[t]
\scriptsize
\centering
\caption{DetaILD of compared language models.}
\vspace{-0.1in}
\label{tab:detaillm}
\begin{tabular}{ll}
\toprule
\textbf{Model} & \textbf{Pre-trained Weights (From Hugging Face)} \\
\midrule
BERT-Base  & \texttt{google-bert/bert-base-uncased}  \\
BERT-Large & \texttt{google-bert/bert-large-uncased} \\
BART-Base  & \texttt{facebook/bart-base}             \\
BART-Base  & \texttt{facebook/bart-large}            \\
RoBERTa-Base  & \texttt{FacebookAI/roberta-base}      \\
RoBERTa-Large & \texttt{FacebookAI/roberta-large}    \\
SimCSE-Base  & \texttt{princeton-nlp/sup-simcse-roberta-base}   \\
SimCSE-Large & \texttt{princeton-nlp/sup-simcse-roberta-large} \\
BLaIR-Base  & \texttt{hyp1231/blair-roberta-base}   \\
BLaIR-Large & \texttt{hyp1231/blair-roberta-large} \\
GTR-Base  & \texttt{sentence-transformers/gtr-t5-base}  \\
GTR-Large & \texttt{sentence-transformers/gtr-t5-large} \\
BGE-Base  & \texttt{BAAI/bge-base-en-v1.5}  \\
BGE-Large & \texttt{BAAI/bge-large-en-v1.5} \\
\bottomrule
\vspace{-0.2in}
\end{tabular}
\end{table}

\subsection{\textbf{Baseline Models}} In this section, we provide a detailed description of the language models compared in this work.

\noindent \textbf{(i) General Language Models.}\vspace{-0.05in}
\begin{itemize}[leftmargin=*]
\item \textbf{BERT}~\cite{BERT}: A prominent transformer model known for its strong language understanding via bidirectional training. We use the pooled BERT output for text embedding. \vspace{-0.1in}
\item \textbf{RoBERTa}~\cite{RoBERTa}: An optimized BERT with dynamic masking and larger datasets. We use the final [CLS] token embedding.\vspace{-0.1in}
\item \textbf{BART}~\cite{BART}: A denoising autoencoder transformer trained on corrupted text reconstruction. We apply mean pooling on the last hidden state for the text embedding.
\end{itemize}

\noindent \textbf{(ii) Language Models for Dense Retrieval.}\vspace{-0.05in}
\begin{itemize}[leftmargin=*]
    \item \textbf{SimCSE}~\cite{SimCSE}: A framework that leverages contrastive learning to generate high-quality sentence embeddings, enhancing the model's ability to discern semantic similarity between sentences.\vspace{-0.1in}
    \item \textbf{GTR}~\cite{GTR}: GTR is a generalizable T5-based dense retriever that improves retrieval tasks across various domains by overcoming limitations of traditional dual encoders.\vspace{-0.1in}
    \item \textbf{BGE}~\cite{BGE}: A state-of-the-art family of well-trained models for general text embeddings, utilizing the English version of BGE.\vspace{-0.05in}
\end{itemize}

\noindent \textbf{(iii) Langauge Models for Recommendation.}\vspace{-0.05in}
\begin{itemize}[leftmargin=*]
    \item \textbf{BLaIR}~\cite{BLaIR}: A series of embedding models for recommendation. BLaIR learns correlations between item metadata and user feedback, improving item retrieval and recommendation.
\end{itemize}

The pre-trained weights utilized for each baseline language models are listed in Table~\ref{tab:detaillm}. For proprietary embedding models \textbf{OpenAIv3}, we use the latest \texttt{text-embedding-3-small} and \texttt{text-embedding-3-large} from OpenAI.

\subsection{\textbf{Detail of Baseline Settings}} Given that the experiment was conducted in a zero-shot setting, where the test data remained unseen during training for both our \model\ and the other baselines, we directly utilized the original released parameters of these baselines from Hugging Face for comparison. Besides, for BGE, we added the recommended retrieval instruction in front of the user profiles, following the provided guidelines from the open-source code, as we found it offered better performance in zero-shot recommendations. We normalized model outputs and computed cosine similarity between user and item embeddings.

\end{appendices}

\end{document}